\documentclass[pre,twocolumn,floats]{revtex4}

\usepackage[latin1]{inputenc}
\usepackage{amsmath}
\usepackage{amsfonts}
\usepackage{epsfig}
\usepackage{bbm}

\begin{document}

\title{Influence of Spatial Correlations on the Lasing Threshold of
Random Lasers}

\author{Michael Patra}
\affiliation{Laboratory for Computational Engineering, 
Helsinki University of Technology, P.\,O. Box 9203, 02015 HUT, Finland}

\begin{abstract}
The lasing threshold of a random laser is computed numerically
from a generic model. It is shown that spatial correlations of the 
disorder in the medium (i.\,e., dielectric constant) lead to an increase
of the 
decay rates of the eigenmodes and of the lasing threshold. This
is in conflict with predictions that such correlations should lower the
threshold.
While all results are derived for photonic systems, the computed decay rate
distributions also apply to electronic systems.
\end{abstract}


\maketitle

In the theory of disordered media two important regimes, diffusive and
localised, are distinguished~\cite{beenakker:97a}. In the diffusive regime (for
weak or moderate disorder), eigenstates are extended and efficient transport is
possible. In the localised regime (for strong disorder), eigenstates become
localised and transport is strongly inhibited. Many experimental findings for
random lasers are more consistent with the assumption of a lasing mode that is
localised while a direct experimental analysis of the sample shows that it is
in the diffusive regime.

To determine whether a sample is in the localised or in the diffusive regime, a
transport property is measured. The most efficient way to achieve this  is
to check for the rounding of the backscattering cone~\cite{schuurmans:99a}.
Such a rounding is not reported from
experiments~\cite{cao:99acao:99bcao:00a,soest:01a}. Transport is, however,
dominated by extend eigenstates, and the simultaneous existence of a few
localised eigenstates in a sample in the diffusive regime, i.\,e., that is
\emph{on average} diffusive, would not be noticed~\cite{burin:01a}. Such
localised modes have recently been detected experimentally in a diffusive
sample~\cite{cao:02a}.

The important question is to explain under which conditions such localised
eigenstates can exist in a diffusive sample. [These states have been termed
anomalously localised states (ALS) or prelocalised states. For a recent review
see Ref.~\onlinecite{mirlin:00a}.] One-dimensional disordered systems are
always in the localised regime, i.\,e., they can never show diffusive
behaviour. Theoretical studies on such systems thus cannot give information on
the interplay between extended and localised modes. The situation is different
in two- and three-dimensional samples. Two-dimensional samples shorter than the
localisation length behave similarly to three-dimensional samples, and one is
allowed to replace three-dimensional systems by their computationally cheaper
two-dimensional counterparts.

The computational cost of treating a two-dimensional sample is
significantly higher than for a one-dimensional sample, and only few
studies have been published~\cite{footnote1}. Reference~\onlinecite{burin:01a}
models the scatterers in the disordered media as dipoles,
Ref.~\onlinecite{sebbah:02a} studies circular particles using the
finite-difference time-domain (FDTD) method. Both publications do not state
explicitly whether their samples are in the diffusive or in the localised
regime but the parameters given strongly suggest that the samples are
in the localised regime.

The only publication so far on the interplay between diffusive and localised
eigenstates inside a diffusive sample seems to be
Ref.~\onlinecite{apalkov:02a}. There it was estimated that localised states
become exponentially more frequent when the disorder inside the sample is
spatially correlated. Motivated by the picture of photons travelling in  closed
loop inside a ring-shape structure~\cite{cao:99acao:99bcao:00a}, they study a
ring-shaped area of higher dielectric
constant. This is a very special situation, and it is not obvious how
characteristic such a special situation is for the entire behaviour. (It should
be noted that the opposite effect, namely that in a localised sample a few
modes become extended when spatial correlations in the disorder are introduced,
is well-understood, see e.\,g. Ref.~\onlinecite{izrailev:99a}.)

In this paper we will study this problem from a more generic approach. The
lasing threshold of a sample is determined by the decay rates of the
eigenstates of the system since the loss (=decay) of photons in the mode has to
be compensated by pumping if the sample is to start lasing action. Following the
approach of Ref.~\onlinecite{patra:03a} we numerically compute the decay rate
distribution of a two-dimensional sample on a suitable grid. (Earlier work on
the lasing threshold of chaotic
cavities~\cite{patra:00afrahm:00aschomerus:00a} cannot be applied
since by construction all eigenstates are
extended~\cite{beenakker:97a}.) We improve on previous work by including
spatial correlations.

We use the Anderson Hamiltonian which describes the motion of an uncharged
particle in a spatially varying potential. The results can directly be applied
also to
photonic systems since the Helmholtz equation with a spatially varying
dielectric constant has the same form. The sample is discretised with lattice
spacing $\Delta$, where for electronic systems $\Delta=\pi/k_{\mathrm{F}}$
($k_{\mathrm{F}}$ is the wave vector at the Fermi level) and for photonic
systems $\Delta=2\lambda/\pi$ ($\lambda$ is the wave length of the light).
This is a natural choice in that there is then one 
propagating mode per transversal lattice
point, and the width of the sample is best measured in terms of the number $N$
of propagating modes.

Transport is modelled by
nearest-neighbour hopping with rate $1$. (The results are easily adapted to
arbitrary speed $c$ of transport.) With a spatially
varying potential $P(x,y)$ the Hamiltonian 
for a sample of length $L=\tilde{L}\Delta$
becomes~\cite{patra:03a}
\begin{multline}
\mathcal{H}_{(x,y),(x',y')} = \delta_{x x'} \delta_{y y'} \left[
		P(x,y) - i ( \delta_{1 y} + \delta_{L y} ) \right] \\
	+ \delta_{y y'} \left( \delta_{x+1,x'} + \delta_{x-1,x'} \right)
	+ \delta_{x,x'} \left( \delta_{y+1,y'} + \delta_{y-1,y'} \right) \;,
	\label{eqHamiltonian}
\end{multline}
with $x=1,\ldots,\tilde{L}$ and $y=1,\ldots,N$. 
The imaginary part of $\mathcal{H}$ models 
coupling of the sample to the outside where we
assume that we operate at the centre of the conduction band.

The spatial correlations are assumed to fall of exponentially such that
$P(x,y)$ takes on random values, normal-distributed with zero mean
and correlator
\begin{equation}
	\langle P(\vec{r}) P(\vec{r}\,') \rangle = w^2 \exp\Bigl( - 
		\frac{|\vec{r}-\vec{r}\,'|}{R_{\mathrm{c}}} \Bigr) \;.
	\label{eqRandom}
\end{equation}
$w$ measures the strength of the disorder, and $R_{\mathrm{c}}$ is
the correlation radius. Since we need to generate a large number of mutually
correlated random numbers, a Fourier based method has to be
employed~\cite{kozintsev:99a}.

The eigenvalues of the matrix $\mathcal{H}$ correspond to the
\mbox{(quasi-)}\discretionary{}{}{}eigenmodes of the system. Their real part
$\omega$ gives the energy (or, for photonic systems, the frequency) of the
mode, and their imaginary part $\gamma$ the decay rate~\cite{footnote2}. We
thus have an eigenvalue problem of a non-Hermitian complex symmetric matrix 
but an eigensolver specifically adopted to this structure
exists~\cite{patra:03a}. Even with this efficient
eigensolver, this is still a numerically expensive task, and it is impossible
to analyse so many samples that there would be no more noise in the results.

\begin{figure}
\centering
\epsfig{file=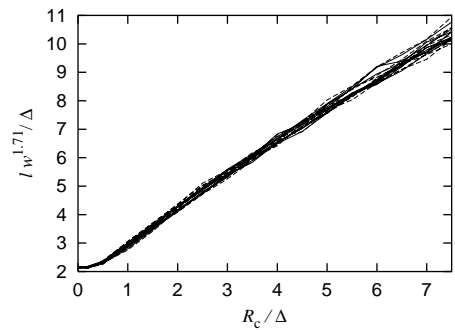,width=2.5in}
\caption{Numerically computed rescaled mean-free path $l$, depending on
the disorder strength $w$ and the correlation radius $R_{\mathrm{c}}$
(in units of the lattice
spacing $\Delta$). The solid lines are for samples of width $N=51$, the dashed
lines for samples of width $N=81$. Samples were computed with $w$ in steps of
$0.1$ and $R_{\mathrm{c}}$ in steps of $0.5\,\Delta$ (plus the value
$R_{\mathrm{c}}=0.2\,\Delta$). By rescaling $l\to l w^{1.71}$ we can demonstrate
the apparent scaling $l\propto w^{-1.71}$ and the factorisation 
$l(w,R_{\mathrm{c}})=f_1(w) f_2(R_{\mathrm{c}})$.}
\label{figMeanfree}
\end{figure}

While the model is described in terms of the disorder strength $w$, contact
with experiments or analytical theories is best made by introduction of the
mean-free path $l$. It can be computed from the length-dependence of the
transmission probability $T$ through the sample. In the diffusive regime,
$l\lesssim L\ll N l$, it is given by~\cite{beenakker:97a}
\begin{equation}
	\frac{1}{T} = 1 + \frac{L}{l} \;.
\end{equation}
The transmission probability has been computed using the method of recursive
Green's functions~\cite{baranger:91a} for variable disorder strength $w$ and
correlation length $R_{\mathrm{c}}$. We determined the mean-free path by
fitting the numerically computed $T(L)$
to this functional form self-consistently in the interval $[l;10\,l]$. (Picking
some other interval, e.\,g. $[0;l]$, changed the result only by about $1\,\%$.)
The rescaled results are depicted in
Fig.~\ref{figMeanfree} for both $N=51$ and $N=81$, i.\,e., for samples of
different width. As the figure shows, both sets of curves are almost identical,
thereby demonstrating that we are operating in the wide-sample regime. The
mean-free path increases significantly as $R_{\mathrm{c}}$ increases. This is
immediately obvious since with increasing $R_{\mathrm{c}}$ the potential changes
less within a given distance; hence, there is less scattering.

We would like to point out two ``curiosities''. The numerical data 
suggest that the mean-free path $l$ factorises as
$l(w,R_{\mathrm{c}})=f_1(w) f_2(R_{\mathrm{c}})$. We did not manage to find an
explanation for this observation. Furthermore, the mean-free path seems to
scale as $l\propto w^{-1.71}$ where $1.71$ is a numerical parameter. 
For uncorrelated random order that is uniformly distributed in the
interval $[-w;w]$, a scaling $l \propto w^{-1.5}$ was found
numerically~\cite{patra:03a}.
An analytical theory is available only for one-dimensional systems in the
limit $w\to 0$ where $l\propto 1/w^2$ is
found~\cite{kramer:93a}, so that a universal scaling for finite $w$ might not
exist at all.

The increase of $l$ with increasing $R_{\mathrm{c}}$ poses a problem for a
systematic study of the effects of correlations: One has to decide whether to
compare samples with identical $l$ (and thus variable $w$) or samples with
identical $w$ (and thus variable $l$). The final results must depend (apart
from trivial prefactors) only on the ratios $L/l$ and $R_{\mathrm{c}}/l$ --- not
on any of those quantities separately. This decision is thus ``only'' one of
numerical efficiency and minimisation of finite-size effects.

For most of our simulations, we have decided to keep $l$ constant at
$l=12.5\,\Delta$. For each value of $R_{\mathrm{c}}$, the needed value for $w$
was determined by interpolation of the numerical data presented 
in Fig.~\ref{figMeanfree}. The choice of constant $l$ offers the  advantage that,
even if $R_{\mathrm{c}}$ is changed, samples with identical ``physical'' length
$L/l$ occupy the same number of lattice points, and thus need the same amount
of numerical work. (For constant ``physical'' length $L/l$, the needed
computing time scales as $\mathcal{O}(l^2)$. With constant $w$, this would
impose severe restrictions on the range of $R_{\mathrm{c}}$ that could be
treated.)

We have computed the decay rates for samples of width $N=50$
for length $L/l=1,2,3,4,5,6,9,12,15,18$, and correlation radius
$R_{\mathrm{c}}/\Delta=0.0,0.2,0.5,1.0,1.5,\ldots,7.5$. For each set of
parameters, approximately 2000 samples were generated. The maximum value of $L$
is limited because we are interested in the diffusive regime, hence
$L$ has to be sufficiently smaller than the
localisation length $L_{\mathrm{loc}}=(N+1)l/2$. We did not consider larger
values of $R_{\mathrm{c}}$ than $7.5\,\Delta$ since the sample should be much
wider than the characteristic length scale
of the disorder. Otherwise the sample
would effectively become one-dimensional.

To check the results, we have computed the decay rate distribution also for
$N=80$ for a few selected values of $L/l$ and $R_{\mathrm{c}}$. To complement
the other simulations, we have kept $w$ constant. As explained above, this
implies that we could only include $R_{\mathrm{c}}\le 2 \Delta$.

Following the approach introduced in Ref.~\onlinecite{patra:03a} for samples in
the diffusive regime, we fit the
numerically computed decay rate distribution to the functional form
\begin{equation}
        \mathcal{P}(\gamma) = \frac{\gamma_0^2}{\gamma^2} 
		\Bigl[ 1 - Q\bigl(M+1,\frac{M \gamma}{\gamma_0} \bigr)\Bigr] \;,
		\label{eqForm1}
\end{equation}
where the fitting parameters $M$ and $\gamma_0$ depend on $N$, $L$ and
$R_{\mathrm{c}}$, and $Q(a,x)\equiv\Gamma(a,x)/\Gamma(a)$ 
is the regularised Gamma function .

All numerically computed histograms fit well to the form~(\ref{eqForm1}). The
dependence of $P(\gamma)$ onto $M$ is only weak for $M\gg 1$, making 
a precise determination of $M$ difficult. Within this error limit, we did
not find a significant dependence of $M$ on $R_{\mathrm{c}}$, and M is
approximately given by the $R_{\mathrm{c}}=0$ result
$M = N / [1 + L/(6 l)]$~\cite{patra:03a}.

\begin{figure}
\centering
\epsfig{file=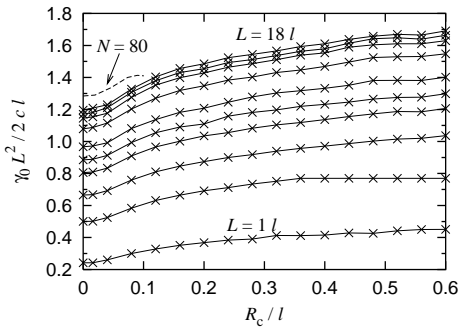,width=2.5in}
\caption{Characteristic decay rate $\gamma_0$ as a function of sample length $L$
and correlation radius $R_{\mathrm{c}}$ (for samples of width $N=50$). The
dashed line is for the control simulations with $N=80$ and $L=25\,l$.}
\label{figGamma}
\end{figure}

The fitting parameter $\gamma_0$, marking the typical value of the decay rates,
can be determined to much better precision. $\gamma_0$ is much more important
for the lasing threshold than $M$, so the limited precision of $M$ does not pose
a problem. The determined $\gamma_0$ is shown in
Fig.~\ref{figGamma}. 

For $R_{\mathrm{c}}=0$, $\gamma_0 L^2$ seems to approach a constant value as
$L$ is increased. This value is about $20\,\%$ larger than the value $1/(2 c
l)$ found  numerically for equi-distributed disorder in the interval
$[-w;w]$~\cite{patra:03a}. Trying to approach the limit $L\to\infty$
numerically is not possible since then the sample  would become localised. 

The important conclusion from Fig.~\ref{figGamma} is that for samples of
arbitrary length, the introduction of correlations in the disorder leads to an
increase of the decay rates. This increase is quick as $R_{\mathrm{c}}$ is
increased starting from $0$, and becomes slower for large $R_{\mathrm{c}}$. The
same behaviour is seen in the control simulations with $N=80$ and fixed $w$
(and thus variable $l$).

Until now, all results are valid for both electronic and photonic systems.
Now we will specialise to random lasers.
The light inside a random laser is amplified by a laser dye. This dye is able
to amplify light within a certain range of frequencies, so only $K\gg 1$ 
eigenmodes out of all eigenmodes of the system are amplified. This number
varies only slightly between different realisations of the same ensemble due to
an effect known as spectral rigidity~\cite{beenakker:97a}. The lasing threshold
is given by the smallest decay rate out of the $K$ modes within the
amplification window~\cite{misirpashaev:98a}. This is immediately obvious since
the lasing threshold is passed when photons are created faster than they can
decay (=escape from the sample).

\begin{figure}[b]
\centering
\epsfig{file=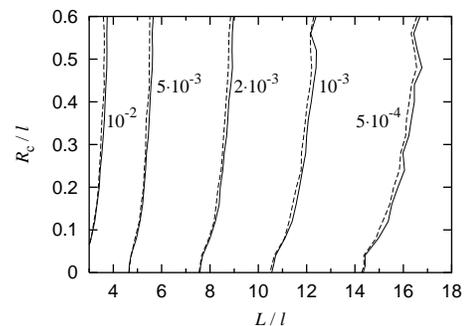,width=2.5in}
\caption{Comparison of the average of the lasing threshold, directly computed 
from the numerical data (solid line), and the most likely lasing threshold
computed from the distribution of the individual decay rates (dashed line).
Both lines have been computed from
the same samples, explaining the correlation of the noise in the two sets of
lines.}
\label{correlfig}
\end{figure}

There are two different approaches to computing the lasing threshold of a
random laser. The direct approach is to compute the eigenmodes of a certain number
of realisations of the disordered systems, then for each realisation to
determine the smallest decay rate inside the amplification window, and finally
collect statistics for those values. Since this process yields only a single
datum per sample, a very large number of realisations needs to be computed to
arrive at data of sufficient quality. The average lasing threshold determined in
this way is depicted in Fig.~\ref{correlfig} as dashed line.

Frequently more efficient is the second approach where one starts with the
computation of the distribution $P(\gamma)$ of the individual decay rates. The
intermediary result is either a numerical histogram, or, by fitting the
histogram to an analytical form, a distribution function that can be evaluated
directly for arbitrary argument. We adopt the latter and use 
Eq.~(\ref{eqForm1})) together with 
the values of $M$ and $\gamma_0$ computed by fitting. 

The distribution $P_{\mathrm{l}}(\gamma_{\mathrm{l}})$ of the lasing threshold
is the distribution of the smallest value out of $K$ values, each distributed
according to $P(\gamma)$. This assumes that the decay rates of different modes
are uncorrelated. For $K\gg 1$ this seems logical but to our best knowledge no
explicit check of this assumption has been published so far. As a side-effect
of our computations, we will fill this gap.

$P_{\mathrm{l}}(\gamma_{\mathrm{l}})$ is difficult to
evaluate numerically for $K\gg 1$ since it is sharply peaked. The position
$\gamma_{\mathrm{m}}$ of the maximum of $P_{\mathrm{l}}$ is immediately seen to
be given by
\begin{equation}
        0 = \frac{d P(\gamma_{\mathrm{m}})}{d\gamma_{\mathrm{m}}} 
                \left[ 1-\int_0^{\gamma_{\mathrm{m}}} P(\gamma') d\gamma' 
                \right]
                - (K-1) [ P (\gamma_{\mathrm{m}})]^2 \;.
        \label{eqtreshold2}
\end{equation}

Since $P_{\mathrm{l}}$ is that sharply peaked, $\gamma_{\mathrm{m}}$ already
contains all the relevant information, and nothing relevant is lost by not
computing the entire distribution. The lasing threshold computed from
Eq.~(\ref{eqtreshold2}), after inserting the fitting parameters $M$ and
$\gamma_0$ computed from the numerical histograms into Eq.~(\ref{eqForm1}), is
shown as solid line in Fig.~\ref{correlfig}.

From the figure, two important conclusions can be drawn. First, the lasing
threshold computed via the two separate methods agrees well. (The ``noise''
of the two sets of curves is correlated since the same raw
data was used as input for both methods.) This means that the decay rates of
different modes indeed are uncorrelated. Furthermore, also fitting the numerical
histogram to the form~(\ref{eqForm1}) is a valid procedure.

The second conclusion --- the heart of this paper --- is that introducing
spatial correlations into the disorder of a random laser increases the lasing
threshold, in contradiction to predictions~\cite{apalkov:02a}.

In this paper, we have thus arrived at two related --- but not identical ---
results. We have shown that the decay rates increase if
spatial correlations of the disorder are introduced. The computed decay rate
distribution possesses the same form, just with different parameter, as
earlier observed for diffusive
samples with uncorrelated disorder~\cite{patra:03a}.
This first result means that the ``typical'' eigenstates become more lossy.

Our second result is that also the lasing threshold increases. This means that
also the ``special'' eigenstates with lower-than-average loss, which are
selected by mode competition to become the lasing modes, become more lossy.
Even though we did not directly compute the spatial extend of the
eigenstates, this still
clearly demonstrates that no (pre-)localised eigenstates are formed by the
introduction of spatial disorder. We thus fail to observe the prediction that
such states should be created~\cite{apalkov:02a}. 

There are several explanations for the difference between our results and
Ref.~\onlinecite{apalkov:02a}. One explanation is that a single
ring-shaped area of increased dielectric constant does lead to the formation 
of a localised state, as suggested by the authors, but the influence of the
disorder around that ring-shaped area significantly reduces this effect.
Another, equally likely, explanation is that in our simulations we are only
able to treat samples of finite size, with a finite number of eigenstates. The
creation of a localised state may be an event that is so rare the we fail to
see such an event occur in our finite-size simulations. On the other hand, the
typical length scale is given by the area per lasing mode, measured
experimentally to be a few $10~\mu\mathrm{m}^2$ in two-dimensional ZnO
films~\cite{ling:01a}, and our samples are larger than this.

To give a definite answer on whether spatial correlations can explain the
formation of localised states, more numerical studies are needed, preferably
using different methods. Specialised but numerically efficient
models~\cite{burin:01a} cannot incorporate
spatial correlations of the dielectric constant.
Two-dimensional FDTD simulations have already been used to describe random
lasers~\cite{sebbah:02a}. They need to make only minimal assumptions and they
can be extended to include arbitrary spatial correlations. FDTD simulations
thus might be a good candidate but diffusive samples need to be larger
than the localised samples studied so far. Given that FDTD is
computationally very expensive, it is not obvious to us whether
this would still be numerically feasible.

This work was supported by the European Union Marie Curie
fellowship HPMF-CT-2002-01794.


\begin{thebibliography}{21}
\bibitem{footnote1}
There has been a large number of studies
using the diffusion approximation. Using this approximation is numerically very
efficient but forbids the
existence of localised states. These studies thus cannot explain the interplay
between extended and localised eigenstates.
\bibitem{footnote2}
Actually, the
imaginary part of the eigenvalue differs by a factor $2$ from the definition of
the decay rate. The same factor $2$ appears in the definition of the lasing
threshold such that those two factors cancel. For ease of writing, we will
in the following refer to $\gamma$ as decay rate, not each time 
mentioning the factor $2$.
\end{thebibliography}

\end{document}